# Growth of PbTe nanowires by Molecular Beam Epitaxy

Sander G. Schellingerhout[1], Eline J. de Jong[1], Maksim Gomanko[2], Xin Guan[1], Yifan Jiang[2], Max S.M. Hoskam[1], Sebastian Koelling[3], Oussama Moutanabbir[3], Marcel A. Verheijen[1], Sergey M. Frolov[2], Erik P.A.M. Bakkers[1]

Affiliations

[1]*Department of Applied Physics, Eindhoven University of Technology, 5600MB, Eindhoven, the Netherlands*

[2]*Department of Physics and Astronomy, University of Pittsburgh, Pittsburgh, Pennsylvania 15260, USA*

[3]*Department of Engineering Physics, École Polytechnique de Montréal, C.P. 6079, Succ. Centre-Ville, Montréal, Québec, Canada H3C 3A7*

**Abstract**

Advances in quantum technology may come from the discovery of new materials systems that improve the performance or allow for new functionality in electronic devices. Lead telluride (PbTe) is a member of the group IV-VI materials family that has significant untapped potential for exploration. Due to its high electron mobility, strong spin-orbit coupling and ultrahigh dielectric constant it can host few-electron quantum dots and ballistic quantum wires with opportunities for control of electron spins and other quantum degrees of freedom. Here, we report the fabrication of PbTe nanowires by molecular beam epitaxy. We achieve defect-free single crystalline PbTe with large aspect ratios up to 50 suitable for quantum devices. Furthermore, by fabricating a single nanowire field effect transistor, we attain bipolar transport, extract the bandgap and observe Fabry-Pérot oscillations of conductance, a signature of quasiballistic transmission.

## Introduction

Semiconductor nanowires (NWs) are a widely studied platform for quantum transport devices[1] due to the quasi-1D confinement that stems from the small radius (<100nm) and high aspect ratio. By combining a semiconductor NW with a superconductor, a topological superconductor can be realized in which Majorana Zero Modes are expected[2–5]. By defining quantum dots, e.g. using gate potentials, spin qubits can be studied, while in NW Josephson junctions gate-tunable transmon qubits were realized[6]. A substantial fraction of the research in the field of NW quantum computation is focused on III-V semiconductors, such as InSb and InAs[2–5]. Group IV nanowires, such as Ge/Si core-shell wires, were recently used to demonstrate ultrafast spin qubit control[7]. However, it has been challenging to integrate them flexibly with superconductors such as Al due to strong interfacial reactions[8–10]. Other issues include disorder and the presence of nuclear spins, creating obstacles for various qubit realizations.

At the same time, group IV-VI materials are relatively unexplored, despite their potential and attractive characteristics[11,12]. Here, we focus on PbTe NWs which as we show form single crystal, defect free NWs. PbTe has large g-factors, a high bulk carrier mobility[13] of over $10^6$ $cm^2/Vs$ and strong spin-orbit coupling. Most importantly, PbTe has an extremely high static dielectric constant of 1400 at 4.2K[14], which is about two orders of magnitude higher than that of group III-V or IV semiconductors. Owing to this high dielectric constant, charged defects are effectively screened and contribute less to disorder. In addition, PbTe could be combined with Pb, which has one of the highest critical temperature of the elemental superconductors. Potentially, a pristine PbTe-Pb interface can be created without inducing interface reactions.

The key finding of this paper is that molecular beam epitaxy (MBE) is a powerful method for synthesizing high aspect ratio defect free PbTe NWs. Previously, PbTe NWs have been synthesized by using electrochemical deposition resulting in polycrystalline growth[15,16], by using a hydrothermal process[17] and by chemical vapor techniques resulting in electron mobilities around 1 $cm^2/Vs$[18–21].

In MBE, in contrast, growth takes place in an ultra-pure environment and can be controlled at a sub-monolayer level. Growth of single-crystalline PbTe NWs in MBE has been reported[22,23]. Dziawa et al.[22] have grown PbTe wires by the vapor-liquid-solid (VLS) growth mode using self-organized Au droplets on a GaAs(111)B substrate. Most important challenges to study and exploit the electronic properties of the reported PbTe NWs are (1) their limited length making it difficult to accommodate multiple electrodes which is necessary for device fabrication and measurement; (2) the tapering of the wires leading to a non-flat surface and a non-uniform quantization energy along the NW length. Volobuev et al.[23] have grown PbTe NWs from Bi-catalysts, preventing tapering at the cost of unintentionally doping the NWs with Bi.

In this work, we address the two main challenges of this material system. We investigate the PbTe NWs growth dynamics in MBE and by tuning the growth parameters, we obtain long, untapered wires with a large aspect ratio. The wires are single crystalline, do not contain foreign impurities above the few ppm level, and from low temperature transport experiments we find that electron transport between the contacts is either ballistic or quasiballistic. Thus our PbTe NWs offer a promising platform for more advanced quantum transport experiments and device fabrication.

## Experimental

Au-catalysed PbTe NWs are grown on GaAs(111)B substrates. The substrates are covered by 20nm of $SiN_x$ using plasma-enhanced chemical vapor deposition (PECVD). Sub-100 nm diameter holes are patterned in a 90 nm thick CSAR6200.04 resist layer using electron beam lithography (EBL). The resist is developed for 60s in proprietary CSAR developer, rinsed in isopropyl alcohol (IPA) for 30s. The holes are transferred into the $SiN_x$ layer using 65s of etching with 30:1 BHF. 8 nm of Au is deposited using electron beam evaporation (EBE), followed by lift-off using Baker PRS3000. The result is a regular pattern of holes in the SiNx mask on the GaAs substrate with Au islands inside the holes. The GaAs substrate is indium soldered on a sample holder and degassed in high vacuum at 300˚C for 60min before being introduced to the MBE growth chamber. We use elemental Pb and Te cells to precisely control the Pb/Te fluxes. The individual beam equivalent pressure (BEP) of each cell is measured using a naked Bayard-Alpert gauge which can be moved in front of the sample holder. The growth time is 3 hours and 40 minutes unless noted otherwise. The growth temperature is measured using a commercial kSA BandiT system.

Transmission Electron Microscopy (TEM) studies were performed using a probe-corrected JEOL ARM 200F, operated at 200 kV. The TEM is equipped with a 100 $mm^2$ SDD Centurio Energy Dispersive X-ray Spectroscopy (EDX) detector. The nanowires have been transferred to a holey carbon film by swiping the holey carbon film over the substrate containing the NWs.

Samples for Atom Probe Tomography (APT) were prepared by transferring NWs in a FEI Helios Nanolab 660 DualBeam system using a method described in detail by Koelling et al[24]. APT was carried out on a LEAP 4000X-HR from Cameca with a laser producing picosecond pulses at 355nm at a repetition rate of a 65kHz. For the analyses laser pulse energies between 5 and 10 pJ were used and a voltage of 2.5-3kV was applied. The machine has a single ion detection system with an efficiency of ~35%. The APT data were reconstructed with IVAS 3.8.5a34. Due to the relatively low voltage applied during the measurement[25], a comparatively high ratio of ~25% multi-hit events[26] and the heavy elements analyzed in this work the detection system does not fully perform up to specifications and the lighter Pb ions, that typically arrive at the detector first during a multi-hit event, make up ~ 60% of the measured bulk concentration of the stoichiometric PbTe.

Single NW field effect transistor (FET) devices are fabricated on highly p-doped Si(100) substrates that act as a global back gate, provided with a 285 nm thermal $SiO_2$ layer on top. NWs are transferred from the growth substrate to the device substrate by mechanical transfer and are located using a scanning electron microscope (SEM). To write the contacts, a double layer of PMMA (495 A4 and 950 A2) is spun at 5000 RPM for 60s and baked for 15 minutes at 175°C. Using the SEM images as reference the design is written using EBL. The resist is developed in MIBK:IPA (ratio 1:3) for 60s, and rinsed in IPA for 60s. A descum is done using a short (15s at 50W) oxygen plasma exposure to remove any remaining resist from the exposed areas. The NW native oxide is removed by applying argon milling, after which a Ti/Au (10nm/140nm) layer is deposited *in-situ* via electron beam evaporation. The remaining polymer is removed in acetone, which subsequently lifts off the Ti/Au from unexposed areas. An additional top gate is deposited, as this allows stronger coupling to the NW. Using atomic layer deposition (ALD) at 120°C a 10nm $HfO_2$ dielectric layer is deposited. For the fabrication of top gates, a similar recipe is used. The same PMMA double layer is used with subsequent EBL, development and oxygen descum. A Ti/Au (10/130nm) layer is deposited and lift-off is done overnight in acetone, followed by a rinse in IPA.

## Results and Discussion

We start by investigating the PbTe NW growth dynamics with the aim to obtain high aspect ratio single crystalline NWs using the vapor-liquid-solid (VLS) mechanism. The main experimental parameters are the substrate temperature and the Pb/Te flux ratio. Figures 1a-c show Scanning Electron Microscopy (SEM) images of PbTe NWs grown at 350°C for varying Pb/Te flux ratios. We find PbTe NWs as well as more isotropically grown, faceted crystalline objects. The NW aspect ratio depends on the flux Pb/Te ratio. In case Pb/Te < 1 (Figure 1a), long and thin NWs are grown with a small catalyst droplet on the top. The NWs have grown from the patterned catalysts confirming the VLS growth mechanism, and have a length ranging from 2 to 3 µm, with a diameter of 40 ± 10 nm. A large proportion of the NWs grow perpendicular or near perpendicular to the surface, the long axis being the <100> axis, as will be discussed below. Based on the orientations of the side facets, we can conclude there is no fixed crystallographic orientation orthogonal to the long axis, indicating that the NWs do not have an epitaxial relationship to the substrate. Increasing the Pb flux to Pb/Te = 1 (Figure 1b) leads to NWs with a larger catalyst droplet due to the increased incorporation of Pb in the Au catalyst and in turn a larger NW diameter of 85 ± 15 nm and a decreased length of 1.4 to 1.8 µm. In Figures 1a and b, most of the NWs and faceted crystals are grown from the patterned holes, showing strong growth selectivity gained by the applied $SiN_x$ mask, although parasitic growth on the mask is not fully suppressed. At Pb/Te > 1 (Figure 1c) the excess Pb leads to NWs with a reduced aspect ratio, a diameter of 105 ± 15 nm and a length of 0.8 to 1.0 µm, as well as a large amount of parasitic growth. We do not observe tapering in any of the as-grown PbTe NWs. The random orientations of the crystals emerging from the holes are attributed to the native oxide on the GaAs substrate, preventing an epitaxial relation between the PbTe and the underlying substrate. Figure 1d summarizes the aspect ratio as a function of the Pb/Te flux ratio, showing a dramatic increase in aspect ratio for decreasing Pb/Te flux ratios. The catalyst droplet size increases with the Pb/Te flux ratio leading to thicker wires. A stable, but smaller aspect ratio is found for Pb/Te > 1, where the catalyst droplet is gradually saturated with Pb and reaches its maximum size. The excess Pb stimulates uncontrolled crystal growth on the substrate surface.

As both Pb and Te have a high vapor pressure it is important to note that the growth temperature strongly affects the quantity of effective adatoms on the sample surface via re-evaporation, which means the total supplied flux and the Pb/Te flux ratio has to be adjusted based on the temperature. The parameter window, substrate temperature versus Pb/Te ratio, for PbTe NW growth is summarized in Figure 1e. The growth conditions that result in Au-catalysed NWs (arbitrarily defined here by an aspect ratio larger than 5) are highlighted in yellow, with black indicating conditions not leading to NW growth. PbTe NW growth has been observed with a substrate temperature between 270 and 360°C and Pb/Te ratios from 0.2 to 1.4. The optimum growth temperature is identified at $T_{opt} = 355 \pm 5$°C at a Pb/Te ratio slightly below 1. It is important to note that at higher Pb/Te flux ratios parasitic growth becomes much more pronounced due to the excess of Pb. For a low Pb/Te flux ratio (0.2), PbTe NW growth is observed at temperatures down to 270°C. In the blue shaded area (Pb/Te > 0.3, T<310°C) parasitic growth dominates over NW growth. Above 370°C (shaded red) the adatom re-evaporation rate becomes too high and the PbTe is unstable, preventing NW growth. The Pb/Te flux ratio tunability facilitates the study of the electronic properties of the PbTe NWs, as the free carrier type (n- or p-type) can be tuned by this ratio resulting in Pb or Te vacancies[27]. It is important to note that although the three different Pb/Te flux ratio regimes presented in Figure 1a-c result in different NW morphologies, all three yield single-crystalline NWs, as presented in SI Figure 1. Finally, the NW growth rate is investigated for the optimum Pb/Te ratio of 0.9 and a growth temperature of 360°C (Figure 1f). The NW length increases linearly with time until a length of approximately 1.8um. When growth time exceeds 3h, the growth rate strongly decreases, indicating that the main driving mechanism of the growth is adatom diffusion over NW sidewalls instead of direct absorption of species into the catalyst droplet. The total NW length is thus limited by the adatoms, which can no longer reach the catalyst droplet before being re-evaporated. We observe longer NWs for a lower Pb/Te flux ratio, which indicates an increased sidewall diffusion length due to the increased Te presence around the NW.

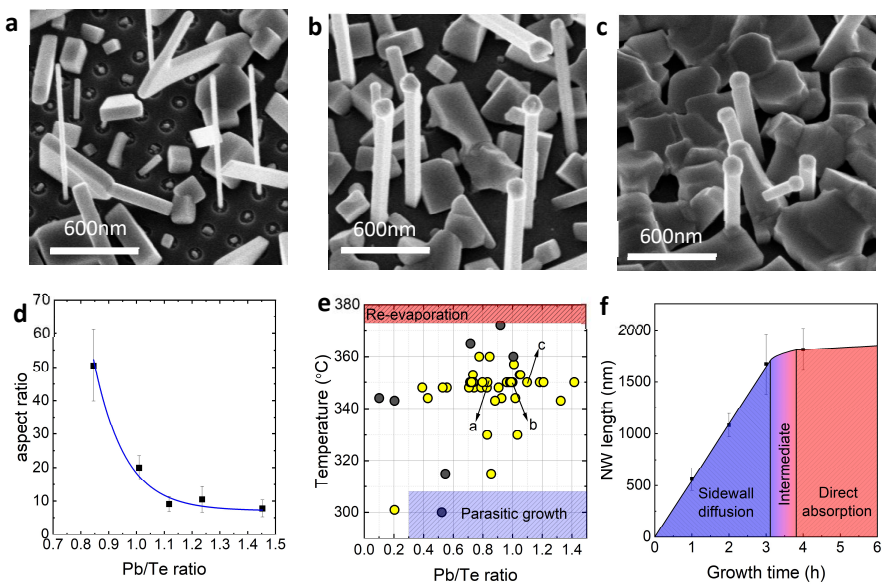

*Figure 1 | PbTe NW growth fundamentals* (a) to (c) 30°-tilted SEM images of representative growth results with constant Te beam flux of 1.24E-7 mbar for a Pb/Te flux ratio of (a) 0.84, (b) 1.0 and (c) 1.1. The growth temperature is 350°C for these samples. (d) Graph showing the NW aspect ratio for different IV/VI ratios. (e) Map of parameter space (substrate temperature and Pb/Te beam flux ratio). Black indicates growth with no resulting NWs, yellow shows Au-catalysed NWs. Red indicates the region where re-evaporation is dominant, blue indicates the region where parasitic growth on the mask is dominant. (f) Graph showing the NW length as a function of the growth time.

Next, we study the crystal structure and growth direction of the PbTe NWs by TEM, shown in Figure 2. The analysis in Figures 2a, b, and c shows that the NWs grow in a <100> crystal direction. Figure 2a reveals a uniform diameter along the NW confirming the non-tapered growth. A thin (2nm) self-terminating native oxide is present on the NW surface (Figure 2b), leading to some roughening of the smooth side facets which could in the future be avoided by *in-situ* passivation in the MBE system. HAADF-STEM and electron diffraction (ED) images acquired along a zone axis orthogonal to the NW axis (Figure 2b, c) show that the NWs are monocrystalline and defect-free. An analysis along a complete NW (SI Figure 2) confirms the monocrystalline structure. The cross sectional TEM image in Figure 2d illustrates that the wires have a square cross section, and are terminated by {200} facets, which are the lowest free-energy facets in PbTe[28]. Energy dispersive X-ray Spectroscopy (EDX) elemental analysis (Figure 2e) shows that the NW indeed consists of both Pb and Te. The elemental mappings of the catalyst particle display a particle predominantly consisting of Pb, with Au present in the core. There are two likely mechanisms that can result in this separation: 1) During growth the Pb and Au are alloyed, after which segregation occurs during cooling down[29], or 2) the Au particle acts as a collection point for Pb at the start of the growth without alloying, allowing a droplet of Pb to form which then facilitates self-catalysed VLS growth. Except the catalyst particle on top the NW, no Au was detected to be incorporated in the NW above the detection limit of EDX (~1 atomic %).

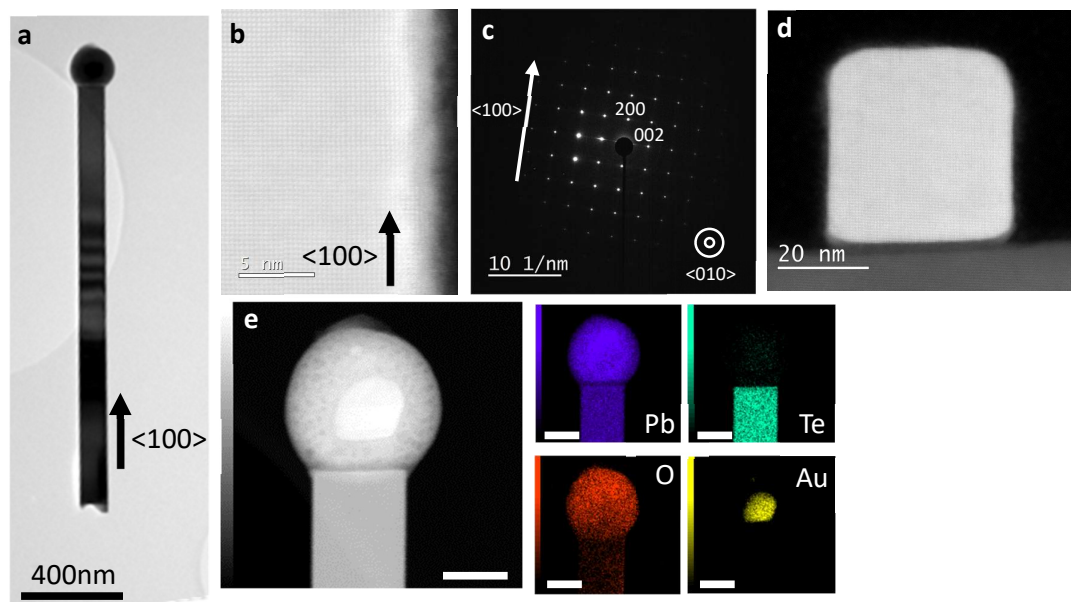

*Figure 2 | TEM of PbTe NWs* (a) Low resolution BFTEM image of a PbTe NW. (b) HAADF-STEM image of a Au-catalysed PbTe NW imaged along the <010> zone axis. The PbTe phase is pure rock salt with no defects. A approx. 2nm native oxide layer is present at the surface. (c) Electron diffraction pattern of the same NW. (d) Cross sectional HAADF-STEM image showing the square cross section. (e) EDX elemental mappings of the top of a Au-catalysed PbTe NW and Au/Pb catalyst droplet. The white line indicates 50nm. The NW consists of pure PbTe, the catalyst is Pb with Au.

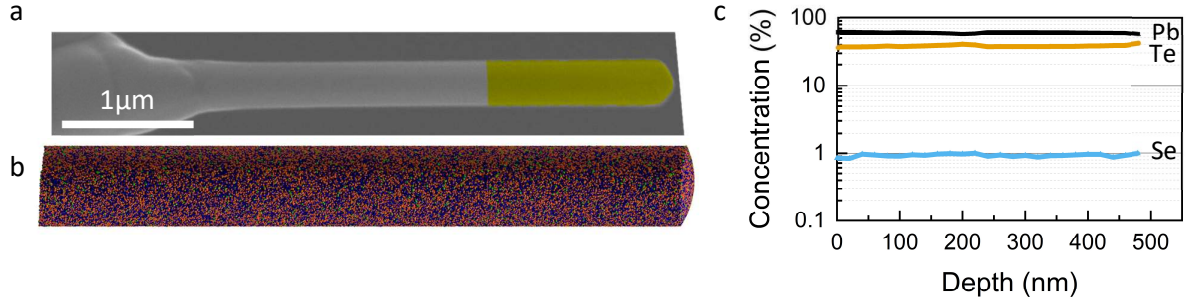

*Figure 3 | APT of a representative PbTe NW. (a) SEM overview image of the NW used for the APT study. The false coloured yellow part is the volume analysed using APT. (b) APT reconstruction of the PbTe NW. Pb is indicated by blue, Te by orange and Se by green. (c) The measured atomic percentages of species present as a function of the axial position in the NW. Slight deviations in the atomic percentage are caused by changes in the tip voltage and laser power.*

The NW composition is further investigated using APT, which has a detection limit of around 1 ppm. An APT sample is prepared following the procedure discussed in the experimental section. An SEM image of a PbTe NW on a post is shown in Figure 3a. The top part, indicated by the yellow area, of this wire has been analysed by APT. A reconstruction of the NW is shown in Figure 3b, indicating a uniform distribution of the elements. The concentration of the detected elements is shown versus the axial position in Figure 3c. Besides Pb and Te, approximately 1 atomic % Se is detected, which is the result of contamination from the Se source in the MBE growth chamber. As Se has the same valency as Te, it will not affect the carrier density in the NWs, but it may have an effect on the carrier mobility due to alloy scattering. No other elements (Au from the catalyst droplet, In from the sample soldering, Ga and As from the growth substrate) are found in the PbTe NW above the detection limit.

In order to study the electronic properties single NW devices are fabricated, shown schematically in Figure 4a. Figure 4b shows an SEM image of a representative device with the distance between the contacts $d_{cont} \approx 300$nm and a NW width $d_{nw} \approx 120$nm. All measurements are done in a dilution refrigerator with a base temperature of 40 mK. The current through the NW is mapped using the source-drain bias $V_{bias}$ and the top gate bias $V_{tg}$ in Figure 4c. The back-gate voltage is kept constant at $V_{bg} = -30$V. The extent of the non-conducting regime reflects the electrostatic doping of the NW within the PbTe band gap, which results in an estimated band gap of $E_g \sim 0.13 \pm 0.03$ eV at 50mK, which is comparable to the expected value[30] of $E_g \sim 0.18$eV at T~0K. The discrepancy may originate from below-gap tunnelling and disorder along the NW. From a linecut at $V_{bias} = 50$mV the typical current as a function of the top-gate voltage is shown. At sufficiently high top-gate voltage ($V_{tg} > -4$V), the Fermi level is tuned into the conduction band and current is mediated by electrons. When decreasing the top-gate voltage, a pinch off regime is present between $V_{tg} = -6.5$V and $V_{tg} = -4$V. When tuning the top-gate voltage even lower ($V_{tg} < -6.5$V), the Fermi level is lowered into the valence band and a hole conductance is measured. The hole current is significantly lower than the electron current, which can be due to a lower hole mobility or a smaller hole transparency at the NW-metal contact[31]. Electron and hole mobilities can, in principle, be estimated from the field effect and from the steepness of the pinch-off traces. The method has limitations, since it requires the unknown gate capacitance as an input, which is challenging to estimate given the ultrahigh dielectric constant of PbTe. A short discussion on the difficulties in estimating mobility is presented in the SI. We obtain indirect evidence of ballistic or quasiballistic transmission from the analysis of quantum oscillations of conductance. In another NW device, using the much weaker coupled back-gate instead of the strongly coupled top-gate, quasi-periodic oscillations are observed in the conductance for a wide range of $V_{bg}$ (Figure 4d). This is a manifestation of Fabry-Pérot type interference in the NW, arising from electrons reflecting either at the NW-metal contacts or between defects within the NW. Likely two or more oscillating modes are present, leading to the slowly changing background of the conductance[32]. Excessive disorder and scattering would lead to the loss of phase coherence and smearing of the Fabry-Pérot oscillations, while their presence can be used to estimate the mean free path and gate capacitance. The length $L_c$ of the segment over which the Fabry-Pérot interference takes place can be approximated by[33] $L_c = \sqrt{\frac{\hbar^2 \pi^2}{2m^* \Delta E}}$, with $m^* = 0.02 m_e$ the effective electron mass[34] and $\Delta E = 0.8 - 1.2$meV the energy spacing of the Fabry-Pérot resonance. This results in $L_c = 125 - 150$nm, which is approximately half of the NW channel and similar to the NW width. This result suggests that the transport in the NW is quasiballistic, with a mean free path in the same length scale as the Fabry-Pérot cavity. Additional data from 3 other NW devices can be found in SI Figure 3. Despite the mentioned uncertainties, this analysis shows that PbTe NWs are promising for quantum devices and should be investigated further as a host of one- and zero-dimensional states.

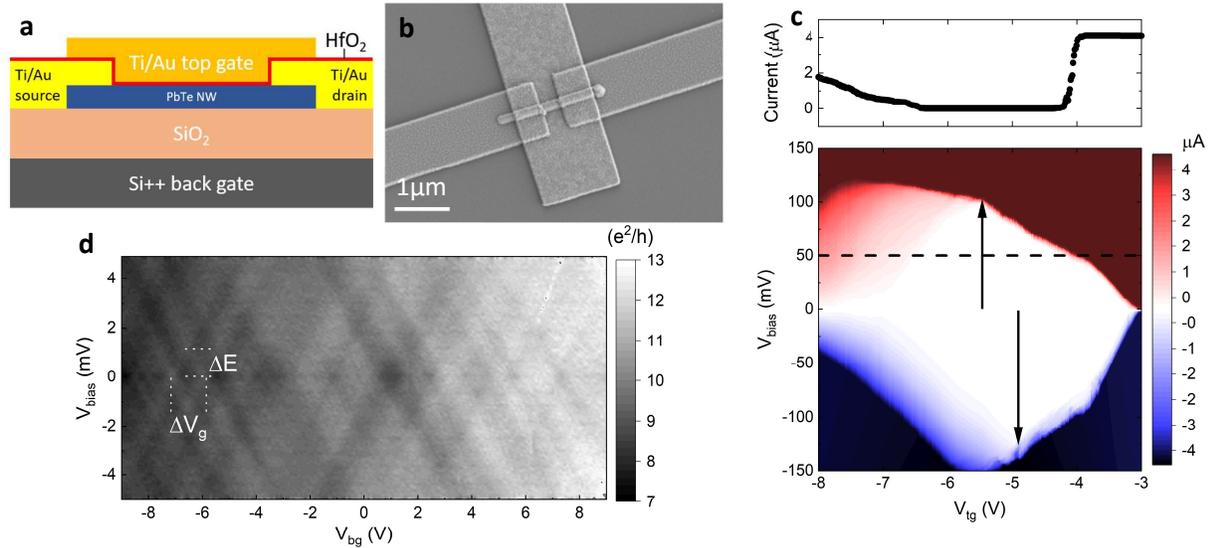

*Figure 4 | FET measurements on a single PbTe NW. (a) Schematic of the PbTe device. (b) SEM image of a representative PbTe device. (c) Current through NW as a function of the bias on the NW and top gate voltage. The back gate voltage is fixed at $V_{bg}$=-30V. The black arrows indicate the extracted band gap. The dashed line indicates the taken linecut at $V_{bias}$=50mV (inset). Clear bipolar behaviour can be seen. (d) Differential conductance as a function of the bias on the NW and the back gate voltage. Clear Fabry-Pérot oscillations are observed, implying (quasi)-ballistic transport in the NW.*

## Conclusions and outlook

Single-crystalline, high aspect-ratio PbTe NWs are grown by the VLS mechanism in an MBE chamber using Au catalysts. The defect-free NWs grow in the <100> direction, with smooth {200} side facets, which are roughened by oxide formation after removal from the vacuum chamber. The Pb/Te ratio during growth determines the NW aspect ratio, and NWs with a length up to 3μm are obtained without tapering. The increased length of the NWs allows for the fabrication of single NW FET devices, in which bipolar transport and Fabry-Pérot oscillations are observed at low temperature. This leads us to believe that electron transport in the device is at least quasiballistic, indicating the high quality of the grown material. Further studies should be directed towards the optimization of the PbTe NWs transport properties. As the oxidized surface likely leads to disorder, the NW surface should be passivated *in-situ* after NW growth or prior to device fabrication. Possible passivation materials are lattice matched crystalline materials with a larger band gap such as CdTe and high-quality dielectrics such as $Al_2O_3$ or $HfO_2$. Additionally, it will be interesting to explore the deposition of superconducting Pb *in-situ* in order to achieve an epitaxial PbTe/Pb semiconductor-superconductor interface.

## Acknowledgements

The authors thank Martijn Dijstelbloem, Marissa Roijen and Bart van der Looij for the support with the MBE reactor and Philipp Leubner and Maarten Kamphuis for helpful discussions. The work in Eindhoven is supported by the European Research Council (ERC TOCINA 834290), and Microsoft Corporation Station Q. We acknowledge Solliance, a solar energy R&D initiative of ECN, TNO, Holst, TU/e, imec and Forschungszentrum Jülich, and the Dutch province of Noord-Brabant for funding the TEM facility. Transport measurements in Pittsburgh are supported by NSF PIRE-1743717, NSF DMR-1906325, ONR and ARO. The APT work was supported by NSERC Canada (Discovery, SPG, and CRD Grants), Canada Research Chairs, Canada Foundation for Innovation, Mitacs, PRIMA Québec, and Defence Canada (Innovation for Defence Excellence and Security, IDEaS).

## Conflict of interest

The authors declare no conflict of interest.

# Supporting information for: Growth of PbTe nanowires by Molecular Beam Epitaxy

Sander G. Schellingerhout[1], Eline J. de Jong[1], Maksim Gomanko[2], Xin Guan[1], Yifan Jiang[2], Max S.M. Hoskam[1], Sebastian Koelling[3], Oussama Moutanabbir[3], Marcel A. Verheijen[1], Sergey M. Frolov[2], Erik P.A.M. Bakkers[1]

Affiliations

[1]*Department of Applied Physics, Eindhoven University of Technology, 5600MB, Eindhoven, the Netherlands*

[2]*Department of Physics and Astronomy, University of Pittsburgh, Pittsburgh, Pennsylvania 15260, USA*

[3]*Department of Engineering Physics, École Polytechnique de Montréal, C.P. 6079, Succ. Centre-Ville, Montréal, Québec, Canada H3C 3A7*

## Section 1. Additional TEM data

As mentioned in the main text, additional TEM data was taken on multiple NWs. SI Figure 1 presents low resolution BFTEM of NWs grown under different Pb/Te ratios (0.84, 1 and 1.1). The electron diffraction patterns indicate that under all three NWs have the same monocrystalline defect-free crystal structure as presented in Figure 2. Additionally, SI Figure 2 presents HRTEM along a full NW grown at a 0.84 ratio. The NW is again monocrystalline and defect-free along its full length.

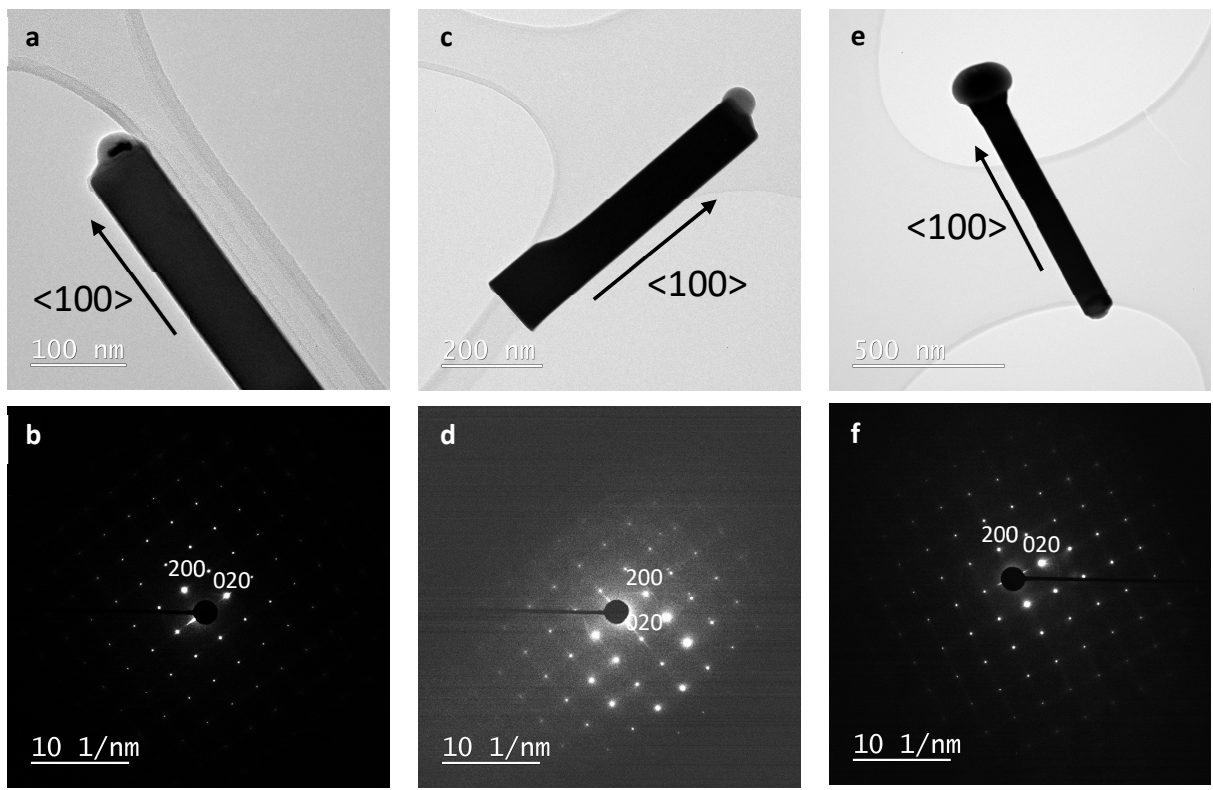

***Supplementary Figure 1 | Defect free NWs for different Pb/Te ratios.*** *Low resolution BFTEM images and electron diffraction patterns for a Pb/Te ratio of (a-b) 0.84, (c-d) 1.0 and (e-f) 1.1. All three NWs have the same defect-free crystal structure.*

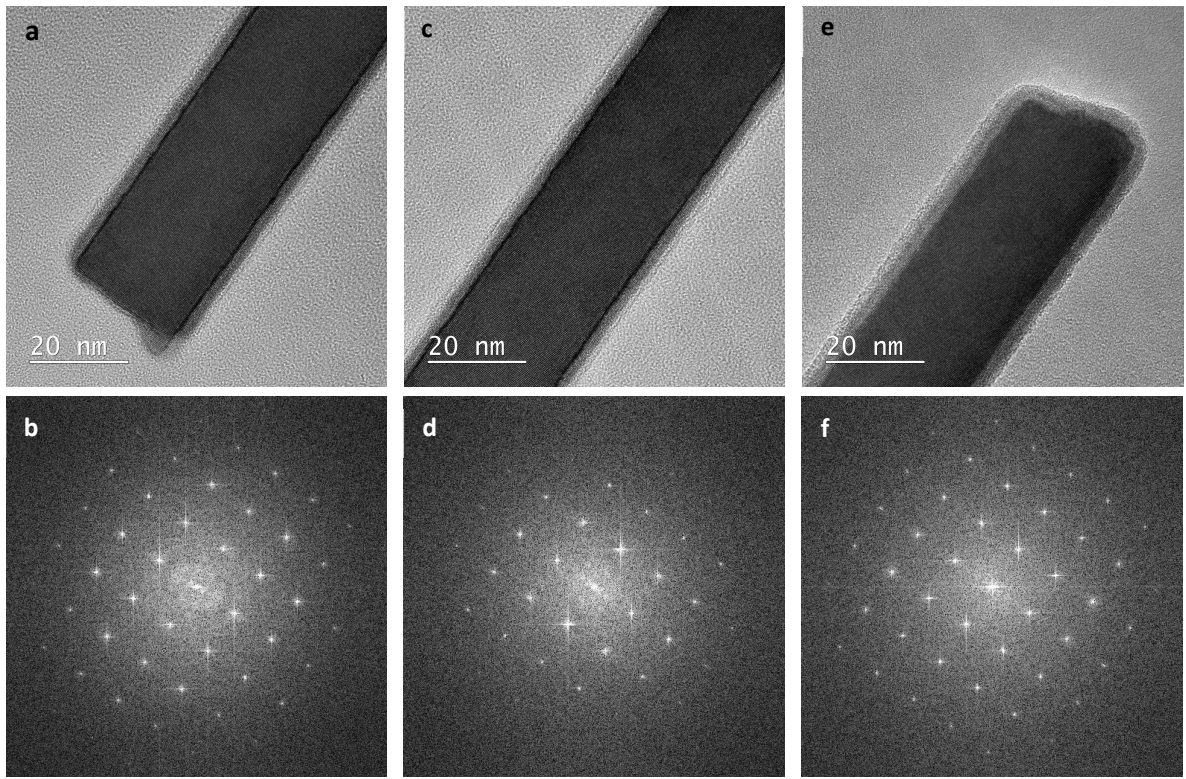

**Supplementary Figure 2 | HRTEM along a full NW.** HRTEM images and FFT patterns along the length of a representative NW. No defects are detected.

## Section 2. Additional Fabry-Pérot data

The conductance oscillations presented in Figure 4 are also measured on 3 additional NW devices. SI Figure 3 shows the greyscale conductance plot of the 3 NW devices with similar dimensions as the device presented in the main text. As in the main text, the length of the Fabry-Pérot cavity is estimated using[1] $L_c = \sqrt{\frac{\hbar^2 \pi^2}{2m^* \Delta E}}$. The dashed white diamonds indicate the area from where the energy spacing $\Delta E$ is estimated. From SI Figure 3a-c we extract respectively $L_c^{(a)} \approx 100$nm, $L_c^{(b)} \approx 100$nm and $L_c^{(c)} \approx 230$nm.

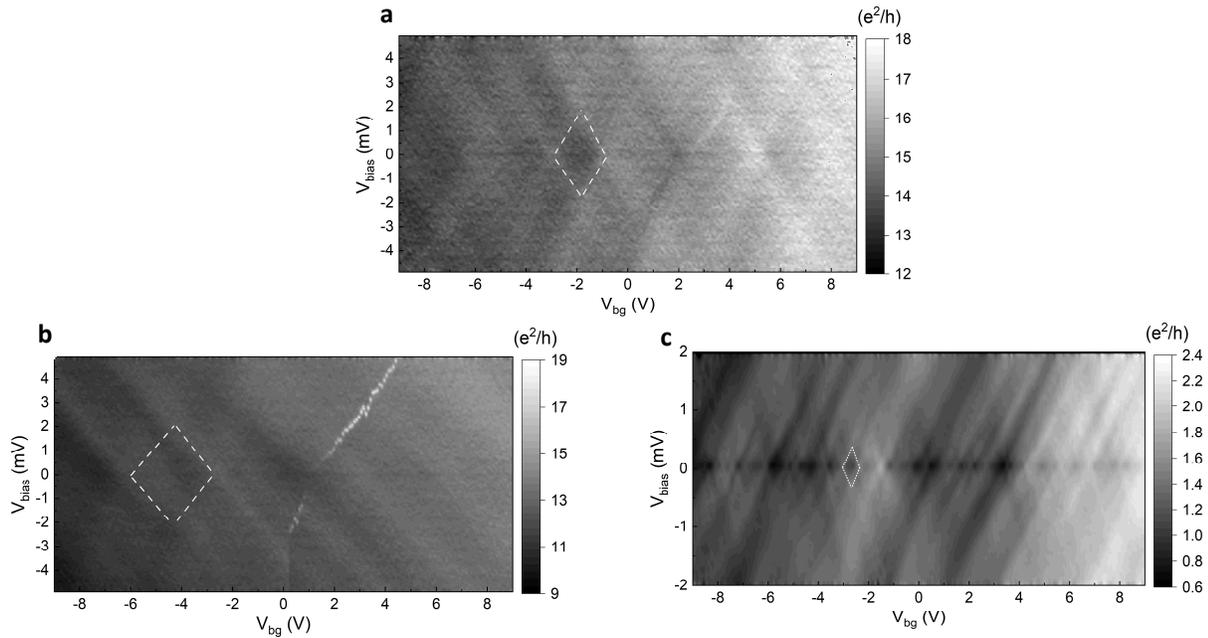

***Supplementary Figure 3 | Fabry-Pérot oscillations.*** *Differential conductance as a function of the bias on the NW and the back gate voltage for 3 different NWs. Clear Fabry-Pérot oscillations are observed, implying (quasi)-ballistic transport. The white dashed diamonds indicate the Fabry-Pérot diamond used to estimate the cavity length.*

**Section 3. Challenges estimating mobility from field-effect pinch-off traces**

Mobility values obtained from pinch-off traces from single NW field-effect devices are prone to a large uncertainty and are often overstated[2,3]. Non-linear transfer characteristics, which is often the case in NW FET devices, can lead to an overestimation of the transconductance and in turn the mobility. On FETs with very short and narrow channels (such as in NW FETs) a contact-limited regime, where the contact resistance is similar or larger than the channel resistance, can lead to nonlinear device characteristics. Additional distortions of the shape of the pinch-off trace can occur due to quantum effects, which can make it steeper due to coulomb effects or shallower due to quantum interference and multiple quantum dots in the NW. Additionally, the calculated mobility is directly dependent on the exact capacitance of the device, which is a similarly uncertain value due to the limited contact spacing which can result in strong screening of the gate electric field, due to the finite density-of-states of the NW[4] and due to the very high dielectric constant of the PbTe itself. Finally, when only a limited gate range can be applied, the saturation resistance is unknown adding additional uncertainty to the mobility extraction. These uncertainties make comparing the apparent mobilities extracted from high mobility field-effect NW devices consisting of different materials (e.g. PbTe to InSb) or device geometries (e.g. top-gate or bottom-gate, channel dimensions) very hard and more than likely misleading.